\documentclass[12pt,preprint]{aastex}

\def\p/{\mbox{$^1$}}
\def\pp/{\mbox{$^2$}}
\def\ppp/{\mbox{$^3$}}
\def\pppp/{\mbox{$^4$}}
\def\m/{\mbox{$^{-1}$}}
\def\mm/{\mbox{$^{-2}$}}
\def\mmm/{\mbox{$^{-3}$}}
\def\mmmm/{\mbox{$^{-4}$}}
\def\Ms/{\mbox{M$_\odot$}}

\def\kms{\mbox{km s$^{-1}$}}

\shorttitle{Pismis 24-1}
\shortauthors{Ma\'{\i}z Apell\'aniz et al.}

\slugcomment{Submitted for publication to the Astrophysical Journal}

\begin{document}

\title{Pismis 24-1: The Stellar Upper Mass Limit Preserved\altaffilmark{1}}

\author{J. Ma\'{\i}z Apell\'aniz\altaffilmark{2,3}}
\affil{Instituto de Astrof\'{\i}sica de Andaluc\'{\i}a-CSIC, Camino bajo de Hu\'etor 50, 18008 Granada, Spain}
\author{Nolan R. Walborn}
\affil{Space Telescope Science Institute, 3700 San Martin Drive, Baltimore, MD 21218, U.S.A.}
\author{N. I. Morrell} 
\affil{Las Campanas Observatory, Observatories of the Carnegie Institution of Washington, Casilla 601, La Serena, Chile}
\author{V. S. Niemela\altaffilmark{4,5}}
\affil{Facultad de Ciencias Astron\'omicas y Geof\'{\i}sicas, Universidad Nacional de La Plata, Paseo del bosque S/N, 1900 La Plata, Argentina}
\author{and E. P. Nelan}
\affil{Space Telescope Science Institute, 3700 San Martin Drive, Baltimore, MD 21218, U.S.A.}


\altaffiltext{1}{This article is based on data gathered with three facilities: the
NASA/ESA \facility{Hubble Space Telescope} (HST), the 6.5 m Magellan Telescopes at 
{Las Campanas Observatory} (LCO), and the 2.15-m J. Sahade telescope at 
\facility{Complejo Astron\'omico El Leoncito} (CASLEO). The HST observations are associated with
GO program 10602. HST is controlled from the Space Telescope Science Institute, which 
is operated by the Association of Universities for Research in Astronomy, Inc., under NASA contract 
NAS~5-26555. CASLEO is operated under agreement between CONICET, SeCyT, and the Universities of La Plata, 
C\'ordoba and San Juan, Argentina.} 
\altaffiltext{2}{e-mail contact: {\tt jmaiz@iaa.es}.}
\altaffiltext{3}{Ram\'on y Cajal fellow.}
\altaffiltext{4}{Member of Carrera del Investigador, CIC-BA, Argentina.}
\altaffiltext{5}{Deceased.}

\begin{abstract}

	Is there a stellar upper mass limit? Recent statistical work seems to indicate that there is and that 
it is in the vicinity of 150 M$_\odot$. In this paper we use HST and ground-based data to investigate the
brightest members of the cluster Pismis 24, one of which (Pismis 24-1) was previously inferred to 
have a mass greater than 200 M$_\odot$, in apparent disagreement with that limit. We determine that Pismis 24-1 
is composed of at least three objects, the resolved \object{Pismis 24-1SW} and the unresolved spectroscopic binary 
\object{Pismis 24-1NE}. The evolutionary zero-age masses of Pismis 24-1SW, the unresolved system Pismis 24-1NE, and
the nearby star \object{Pismis 24-17} are all $\approx 100$ M$_\odot$, very large but under the stellar upper mass limit. 

\end{abstract}


\section{Introduction}

	Massive stars strongly influence the structure and evolution of galaxies; thus, accurate 
knowledge of their fundamental parameters is critical. The spectroscopic binary frequency alone for 
OB stars is near 50\%; moreover, they are preferentially formed in compact multiple systems 
\citep{Masoetal98}.  Multiplicity therefore introduces a basic uncertainty in the determination of 
OB-star parameters, and it must be established with the greatest precision. Radial velocities are 
sensitive to separations up to a few AU; larger separations must be detected by means of high 
angular resolution. Even in the near solar neighborhood, barely resolved or unresolved OB multiple 
systems are known (\citealt{Walbetal99b}; \citealt{Nelaetal04}, and references therein). The 
unobservable gap in separations of course increases with distance, but it will be reduced with more 
powerful instruments.

	One important desideratum is the stellar upper mass limit. The O3 \citep{Walb71b} and O2 
(\citealt{Walbetal02b} [W02]) spectral types include some of the most massive stars known. However, it 
is essential to know their multiplicities in order to determine their actual masses and solve the 
existing discrepancy between the highest accurate mass measured spectroscopically (83 M$_\odot$, 
\citealt{Rauwetal04}; \citealt{Bonaetal04}) and the photometric masses in the range 120-200 M$_\odot$ 
found by several authors. Another important desideratum is knowing how multiplicity influences the 
calculation of the initial mass function (IMF) 
at its high-mass end. Unresolved multiple systems artificially flatten the IMF 
by shifting objects from lower-mass bins into higher-mass ones. A sign of that effect is the change 
in the $\log M-\log L$ slope in the empirical relation derived from binary systems 
\citep{NiemGame04}. At first glance, the Magellanic Clouds would appear to be the ideal 
place to study the flattening of the IMF, given their low extinction and foreground contamination and
the small contribution of depth effects. However, the very massive stars there are located typically 
an order of magnitude farther away than their Galactic counterparts; hence, for the Magellanic Clouds
we are even more limited by the angular resolution of our instruments.

	Several authors \citep{WeidKrou04,Fige05,OeyClar05,Koen06} have recently used the observed 
distribution of stellar masses in R136 and other young clusters and associations to establish that
either (a) an stellar upper mass limit of $\sim$150 M$_\odot$ exists or (b) the IMF becomes much 
steeper beyond 100 M$_\odot$. However, when W02 analyzed the earliest O-type stars
known, they found two apparent exceptions to that limit, the most egregious being Pismis 24-1
(also known as HDE 319718A and LSS 4142A). Two different mass estimates for that star, based on the 
evolutionary calibration of \citet{Vaccetal96} and on the models by \citet{Schaetal92}, yield 
291 M$_\odot$ and 210 M$_\odot$, respectively. Pismis 24-1 was classified as an O4 (f) star by 
\citet{Lortetal84}, as O3 III by \citet{VijaDrill93}, as O3 If* by \citet{Massetal01}, and as 
O3.5 If* by W02. Another star in the same cluster, Pismis 24-17 (also known as HDE 319718B), which also 
appears to be very massive 
(101 M$_\odot$ according to W02), was classified as O3-4 V by \citet{Lortetal84}, as O3 III(f*) by 
\citet{Massetal01}, and as O3.5 III(f*) by W02. Both objects were recently shown by \citet{Morretal05} to 
have the same O\,{\sc iv} and N\,{\sc iv} features in the 3400 \AA\ region as other stars with similar 
spectral classifications, thus confirming their spectral types. Pismis 24-1 and 24-17 are the main
sources of ionizing photons for the H\,{\sc ii} region G353.2+0.9 in NGC 6357 \citep{Bohietal04}.

	In this paper we present new observations that resolve the discrepancy between the
$\sim$150 M$_\odot$ stellar upper mass limit found in clusters and the larger apparent mass of
Pismis 24-1.

\section{Data}

\subsection{Imaging}

\begin{figure}
\begin{center}
\centerline{\includegraphics*[width=0.35\linewidth]{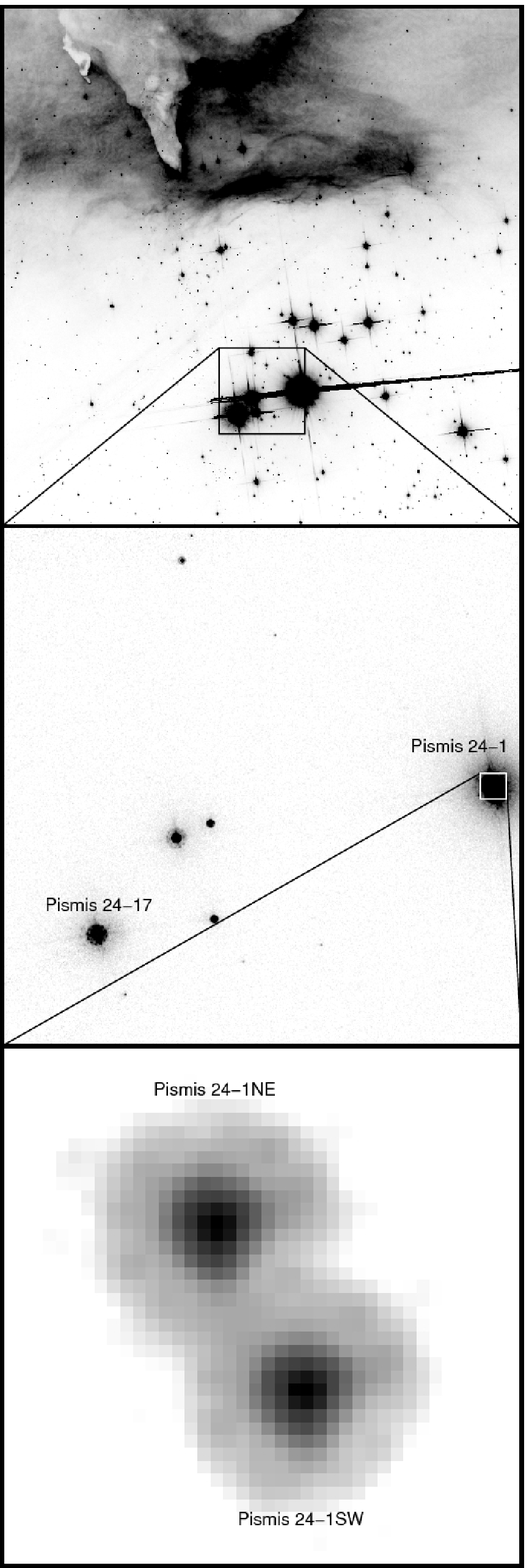}}
\end{center}
\figcaption{HST/ACS imaging of Pismis 24. 
(top) $2\arcmin\times2\arcmin$ FOV from the WFC F850LP drizzled data.  
(middle) $20\arcsec\times20\arcsec$ FOV from the HRC F850LP drizzled data. 
(bottom) $1\arcsec\times1\arcsec$ FOV from the HRC F435W drizzled data. 
In all cases N is towards the top and E towards the left. In the top and middle panels the intensity scale 
is linear while in the bottom one a logarithmic scale was selected to highlight the PSF. The bright stars 
in the top panel are heavily saturated. \label{acsimag}}
\end{figure}

	In order to study the multiplicity at the high-mass end of the IMF, we were granted \facility{HST}
time in GO program 10602. The sample was drawn from an updated version of the Galactic O star catalog
\citep{Maizetal04b} by choosing all stars with spectral types O3.5 and earlier. A total of 20 objects
with previous spectral classifications as O2-3.5 in 12 fields were observed with ACS/WFC (the Wide Field
Channel of the Advanced Camera for Surveys) in F435W+F550M+F850LP and with ACS/HRC (the High Resolution
Channel of the same instrument) in a variety of filters (one of the objects had already been 
observed in another of our programs, GO 10205). The Pismis 24 field was observed on 5 April 2006 using
the above mentioned filters for the WFC and F220W + F250W + F330W + F435W + F550M + F658N + F850LP for the 
HRC (Fig.~\ref{acsimag}). The HRC data were centered on the core of Pismis 24 and were dithered using 2- or 
4-point patterns, depending on the filter. The individual HRC exposure times were selected to maximize 
S/N in all filters for the brightest stars while avoiding saturation. The resulting total integration 
times ranged from 2 s (F850LP) to 704 s (F220W). The WFC data cover a large fraction of the cluster 
and were dithered using a 2-point pattern. The WFC exposures had total integration times of 
460 to 678 s for each of the three filters. No short WFC exposures could be added due to the large 
overheads involved; hence, the bright stars were heavily saturated, as evidenced by the obvious 
bleeding in the top panel of Fig.~\ref{acsimag}.

	The ACS/HRC is an excellent detector for high-resolution imaging, with a pixel size of 
$\approx 0\farcs028$ (the exact value depends on the position), a well-characterized PSF, and a very 
precise geometric distortion solution \citep{AndeKing04,Maiz07b}. Those properties allowed us to easily 
resolve Pismis 24-1 into two components of similar brightness (see bottom panel of Fig.~\ref{acsimag})
in all of the seven HRC filters\footnote{The WFC, with a pixel size approximately double that of the 
HRC, can also resolve the system. The strong saturation prevents this from being immediately obvious in 
the top panel of Fig.~\ref{acsimag}. However, a closer inspection of the bleeding trail that originates 
from Pismis 24-1 reveals that it is double.}. Given the similar magnitudes of the two components (see 
below), we will use the notation Pismis 24-1NE and Pismis 24-1SW in order to avoid possible confusions. 
Pismis 24-17, on the other hand, appears to be unresolved in all of the HRC filters, with its profile 
being essentially indistinguishable (apart from S/N effects) from that of the other bright isolated 
star in the HRC field. 

	One of us (J.M.A.) has recently written a crowded-field photometry package, JMAPHOT, 
especially tailored for the HST instruments. For ACS, the code utilizes the SCI\_COR (scientific,
corrected) files which are the result of applying {\tt multidrizzle} to the FLT (flat-fielded) files in 
order to produce a DRZ (drizzled) frame. The
SCI\_COR files have the advantages of being free of cosmic rays and hot pixels and, at the same time,
having the same geometric properties of the FLT files. Using geometrically uncorrected data allows for 
more precise position measurements and does not modify the photon statistics (i.e. adjacent pixels do
not have correlated errors), thus improving the reliability of PSF fitting \citep{AndeKing04}. On the 
other hand, working on a geometrically distorted frame introduces some inconveniences, such as making 
the calculation of positions and aperture corrections less straightforward. When bright isolated stars 
are present, as is the case for our Pismis 24 data, JMAPHOT can use them to model the PSF residuals and 
the aperture corrections.

	We have applied JMAPHOT to the Pismis 24 HRC data. The results for Pismis 24-1NE, 24-1SW, and 
24-17 are shown in Table~\ref{hrcphot}. The magnitudes have been corrected for CTE (Charge Transfer Efficiency)
and aperture effects and their uncertainties are calculated from the measured S/N plus a contribution from 
those effects. The aperture correction for F850LP requires a special treatment due to its strong dependence 
with the spectral energy distribution or SED \citep{Sirietal05}: we have used a value in accordance with the
large extinction of Pismis 24 and we have also added 0.05 magnitudes in quadrature to the
F850LP uncertainties to account for this effect. Also, as of the time of this writing there is no 
low-order flat field calculated for the F220W and F250W filters of ACS/HRC. For that reason, we have also
added an additional factor of 0.05 magnitudes to the uncertainties for those filters.

\begin{deluxetable}{lccccccc}
\tabletypesize{\scriptsize}
\tablecaption{ACS/HRC photometry expressed in VEGAMAG \label{hrcphot}}
\tablewidth{0pt}
\tablehead{Object &       F220W      &       F250W      &       F330W      &       F435W      &       F550M      &       F658N      &      F850LP     }
\startdata
Pismis 24-1NE     & 16.628$\pm$0.051 & 14.513$\pm$0.051 & 13.103$\pm$0.010 & 12.781$\pm$0.006 & 10.996$\pm$0.017 &  9.846$\pm$0.005 & 8.360$\pm$0.050 \\
Pismis 24-1SW     & 16.741$\pm$0.051 & 14.621$\pm$0.051 & 13.202$\pm$0.010 & 12.897$\pm$0.006 & 11.111$\pm$0.017 & 10.030$\pm$0.005 & 8.510$\pm$0.050 \\
Pismis 24-17      & 17.504$\pm$0.051 & 15.451$\pm$0.051 & 14.027$\pm$0.012 & 13.654$\pm$0.006 & 11.783$\pm$0.023 & 10.640$\pm$0.004 & 9.011$\pm$0.050 \\
\enddata
\end{deluxetable}

	Our HRC data consist of 24 individual exposures. Combining the individual JMAPHOT measurements 
for the separation and position angle with their corresponding uncertainties we measured values of
363.86$\pm$0.22 mas and 207.816$\pm$0.015 degrees E of N, respectively. 

\subsection{Spectroscopy}

\subsubsection{Las Campanas}

	Spectroscopic observations of Pismis 24-1 NE and SW were obtained with the Inamori Magellan Areal 
Camera and Spectrograph (IMACS) on the Magellan I (Baade) telescope at Las Campanas Observatory 
(\facility{LCO}), on May 9.41 UT, under excellent seeing conditions (0.4\arcsec\ - 0.5\arcsec), 
which are characteristic of the Magellan telescopes. We used the f/4 (long) camera with a projected slit 
width of 2.08 \AA\ to cover the wavelength range 3660--6770~\AA\ at a reciprocal dispersion of 
0.38 \AA px$^{-1}$. A spectrum of Pismis 24-17 was obtained a few 
minutes later, with the same instrumentation but somewhat poorer seeing ($\sim$ 0.6\arcsec). A He-Ne-Ar 
comparison spectrum and a flux standard were also observed, as well as the usual series of bias and dome 
flats. These data were processed at LCO, making use of standard IRAF\footnote{IRAF is distributed by the 
National Optical Astronomy Observatories, which are operated by the Association of Universities for Research
in Astronomy, Inc., under cooperative agreement with the National Science Foundation.} routines, to yield two 
dimensional wavelength and flux calibrated images.

\begin{figure}
\begin{center}
\centerline{\includegraphics*[width=0.900\linewidth]{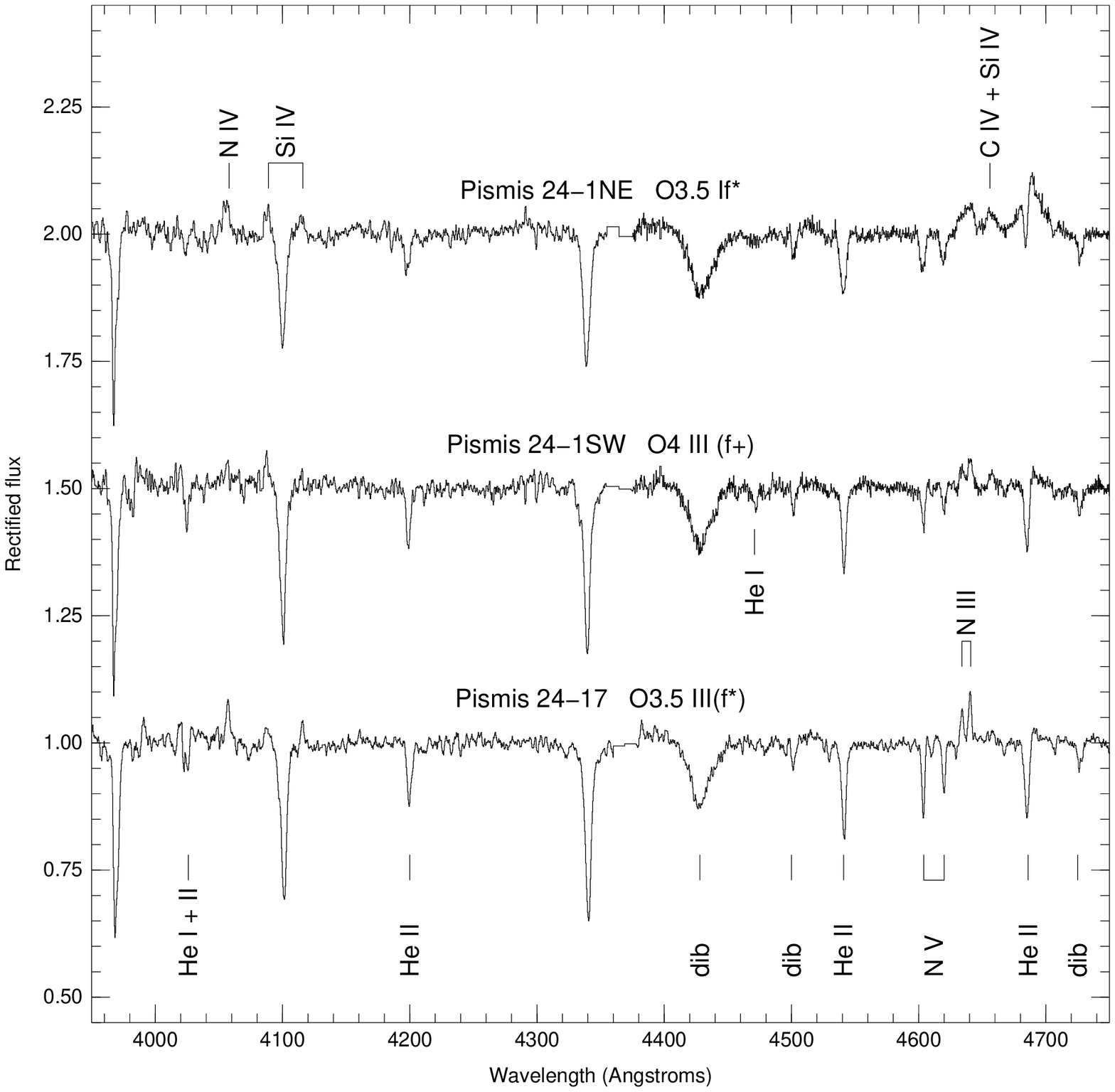}}
\end{center}
\figcaption{Rectified linear-intensity spectrograms of Pismis 24-1NE, 24-1SW, and 24-17 obtained with 
the IMACS-Long Camera of the Baade telescope at Las Campanas Observatory. Each spectrogram has two 
segments, a violet one and a blue-green one, joined at 4370 \AA. The violet segments have been smoothed 
with 3-, 3-, and 5-pixel boxes, respectively (from top to bottom). The blue-green segment for Pismis 
24-17 has been smoothed with a 3-pixel box. Constant values of 0.5 and 1.0 have been added to the
Pismis 24-1SW and 24-1NE spectrograms, respectively. The non-dib lines identified are, in order of
increasing wavelength, He\,{\sc i}+{\sc ii}~$\lambda$4026, N\,{\sc iv}~$\lambda$4058, 
Si\,{\sc iv}~$\lambda\lambda$4089-4116, He\,{\sc ii}~$\lambda$4200, He\,{\sc i}~$\lambda$4471,
He\,{\sc ii}~$\lambda$4541, N\,{\sc v}~$\lambda\lambda$4604-4620, 
N\,{\sc iii}~$\lambda\lambda$4634-4640-4642, Si\,{\sc iv}~$\lambda$4654, C\,{\sc iv}~$\lambda$4658,
and He\,{\sc ii}~$\lambda$4686, with dib standing for diffuse interstellar band. \label{lcospec}}
\end{figure}

	Given the separation of only slightly more than 3 pixels between the two Pismis 24-1
components in the IMACS Pismis 24-1 spectra, we used MULTISPEC \citep{Maiz05a} for the extraction.
That IDL code is designed to obtain spectra from slitless or long-slit exposures of crowded fields using 
PSF fitting along the direction perpendicular to the dispersion. The extracted spectra for the three 
massive stars are shown in Fig.~\ref{lcospec}.

\subsubsection{CASLEO}

 	Since \citet{Lortetal84} had suggested from observed radial velocity variations that Pismis 24-1 
probably is a spectroscopic binary, we acquired high resolution spectra to test this hypothesis.
Spectral images of Pis24-1 were obtained with the 2.15-m J. Sahade telescope at Complejo Astron\'omico
El Leoncito (\facility{CASLEO}) in San Juan, Argentina, during two
observing runs: September 1998 and June 2000. We used the REOSC echelle spectrograph with a TEK 
1024$\times$1024 pixels detector. The spectral resolving power was $R \sim 15000$. We obtained 5 exposures
covering the wavelength region 3800--6100~\AA\ and 3 exposures covering 4400--6700~\AA.  We selected a 
slit width of 2\arcsec\ for all of our REOSC spectra, as this was the average seeing during our 
observations. Exposure times for the stellar images ranged between 30 and 45 minutes, resulting in 
spectra of signal-to-noise ratio $\sim$50. 

 	All the spectral images obtained at REOSC were processed and analyzed with standard IRAF 
routines. We determined radial velocities of the four most conspicuous absorption lines in our 
spectra, namely the hydrogen absorptions H$\gamma$ and H$\beta$, and the He\,{\sc ii} absorptions 4541 
and 5411 \AA, fitting gaussian profiles within the IRAF routine {\it splot}. Also, nebular and 
interstellar lines were measured, and their velocities remained constant within the errors (2-3~\kms.)

\section{The most massive stars in the core of Pismis 24}

\subsection{Results from spectroscopy}

	The spectral type of Pismis~24-1NE derived from the IMACS data, O3.5~If*, 
is consistent with that assigned by W02 to the 
composite observation of the system by Massey et al. (2001).  Indeed, that spectrum defined the new type 
introduced by W02. The key criterion is the comparable intensity of N\,{\sc iv} $\lambda$4058 and
N\,{\sc iii} $\lambda\lambda$4634-4640-4642 (which stretches the original meaning of ``f*'' that the 
N\,{\sc iv} is stronger than the N\,{\sc iii}). Pismis~24-1SW is of later type from both the 
N~\,{\sc iv}/N~\,{\sc iii} ratio (the former is weak but still detected) and 
the weak He\,{\sc i} $\lambda$4471 absorption line. Indeed, it now appears that the latter feature, as 
well as the strength of He\,{\sc i}+{\sc ii} $\lambda$4026, in the composite data is due to the SW 
component. Such is consistent with the nearly identical magnitudes of the two components and the relative 
morphology of their spectra. The spectral type of Pismis~24-1SW derived from the IMACS observation is 
O4~III(f+), where the luminosity class 
follows from the weakened absorption and incipient emission at He\,{\sc ii} $\lambda$4686, and ``f+'' 
means that Si\,{\sc iv} $\lambda\lambda$4089-4116 are in emission in addition to the N\,{\sc iii} (which 
is subsumed in the definition of f*). The spectral type of Pismis~24-17, O3.5~III(f*), is also unchanged 
from the discussion by W02. The overlap between giant and supergiant absolute visual magnitudes at the 
earliest spectral types is discussed by W02; in Pismis~24-1 at least one additional component contributes 
to the relative apparent magnitudes of the resolved pair.

  	Our REOSC spectra were obtained with worse seeing than the IMACS ones and it is obvious that 
they contain the combined spectrum of both visual components, Pismis 24-1NE and Pismis 24-1SW, which are 
less than 0\farcs4 apart. The absorption lines in our spectra appear to have variable shape, but we did 
not observe clearly separated component lines in our spectra. We determined the radial velocities of the 
mean profiles of the absorption lines, with the aim of verifyng the previously published radial velocity 
variations, which most probably would imply that at least one of the visual components of Pis24-1
also is a short period spectroscopic binary. The mean radial velocities of the absorption lines measured 
in our eight spectra present variations from $+$20 to $-$90 \kms\ (heliocentric), thus confirming with 
somewhat larger amplitude the variations published by \citet{Lortetal84}. Since we are measuring the lines 
blended with the visual companion, the actual amplitude of the radial velocity variations may be 
considerably larger. Our data are too sparse to deduce a value for the orbital period, suffice it to say 
that it seems to be of only a few days.

  	Our spectra show the signatures of a turbulent and high temperature interstellar 
medium commonly observed in H\,{\sc ii} regions surrounding very hot stars: namely that the interstellar 
lines have multiple components, and the presence of nebular absorption of He\,{\sc i} 3888\AA . The 
main components of the interstellar absorptions of Na\,{\sc i} D1 and D2 in our spectra have 
heliocentric radial velocities of $-13\pm 2$ \kms\ and $-12\pm 3$ \kms.

	Phil Massey and collaborators (private communication) have 
detected optical variability in the unresolved Pismis 24-1NE+SW photometry with a peak-to-peak
amplitude of 0.07 magnitudes and a period of 2.36088 days. This variability, together with the observed 
radial velocity variations, is a clear sign that Pismis 24-1 is at least a triple system. Furthermore, 
the radial velocity of the N\,{\sc iv}~$\lambda$4058 line in the IMACS spectrum coincides with the value 
derived from our radial velocities obtained from the REOSC spectra at the appropriate orbital 
phase\footnote{Note, however, that we have insufficient points to determine a full velocity curve, which
we will obtain in the future.}. Since N\,{\sc iv}~$\lambda$4058 is visible only in the spectrum of Pismis
24-1NE, it must be the spectroscopic binary. We also point out that the fit of the X-ray spectrum of 
Pismis 24-1 requires a two-temperature plasma model \citep{Wangetal06}, which is usually an indication of
wind-wind interaction in a relatively close binary. The results here point towards Pismis 24-1NE as the
source of the hard X-ray component.

\subsection{Results from photometry}

	The results in Table~\ref{hrcphot} show that Pismis 24-1NE is brighter than 24-1SW in the 
optical/NUV by $\sim 0.1$ magnitudes, with nearly identical colors. Given their very similar spectral 
type and location, this is as expected: both stars have almost the same intrinsic colors and their light is
affected by the same amount and type of dust. The most significant difference is in F550M$-$F658N 
[$V-$(H$\alpha$+continuum)], where Pismis 24-1NE is redder by 0.069$\pm$0.025 magnitudes. This is also 
expected, since 24-1NE is a supergiant and has a stronger H$\alpha$ emission (this effect is also seen
in our LCO spectra).  Pismis 24-1NE is brighter than 24-17 by 0.924$\pm$0.016 magnitudes in F330W but the 
latter is slightly redder than either of the 24-1 components; hence, the difference at F850LP is reduced 
to 0.651$\pm$0.007 magnitudes\footnote{Most of the uncertainty for F850LP is due to the aperture correction,
so it should not be applied to magnitude differences within the same filter.}. The likely source for the 
color differences is that the light from Pismis 24-17 is more extincted than that from 24-1.

\begin{deluxetable}{lcccccrc}
\tabletypesize{\small}
\tablecaption{CHORIZOS results. \label{chotable}}
\tablewidth{0pt}
\tablehead{Object & 2MASS & $A_{\rm F550M}$ & $R_{5495}$    & $M_V$            & $(m-A+BC)_{\rm F550M}$ & $\chi^2_{\rm min}$ & dof}
\startdata
Pismis 24-1NE     & no    & 5.54$\pm$0.10   & 2.87$\pm$0.05 & $-$6.41$\pm$0.17 & 1.71$\pm$0.10          &  9.1               & 1 \\
Pismis 24-1SW     & no    & 5.52$\pm$0.10   & 2.87$\pm$0.05 & $-$6.28$\pm$0.17 & 1.73$\pm$0.10          &  7.0               & 1 \\
Pismis 24-17      & no    & 5.91$\pm$0.11   & 2.94$\pm$0.05 & $-$6.01$\pm$0.17 & 1.91$\pm$0.10          &  4.6               & 1 \\
Pismis 24-1       & yes   & 5.87$\pm$0.02   & 3.01$\pm$0.01 & $-$7.50$\pm$0.14 & 0.57$\pm$0.02          & 40.4               & 4 \\
Pismis 24-17      & yes   & 6.34$\pm$0.02   & 3.11$\pm$0.02 & $-$6.50$\pm$0.14 & 1.42$\pm$0.02          & 26.1               & 4 \\
\enddata
\end{deluxetable}

	We have processed the Pismis 24-1NE, 24-1SW, and 24-17 photometry with CHORIZOS \citep{Maiz04c}, a
Bayesian code that compares observed magnitudes with SED families and calculates the likelihood 
distribution. The current version of the code includes the most recent calibration for optical and NIR 
photometric zero points \citep{Maiz07a}. For each of the stars we have fixed the values of the
temperature and gravity derived from their spectral types and the calibration of \citet{Martetal05} and we
have left $R_{5495}$ and $E(4405-5495)$, the monochromatic equivalents of $R_V$ and $E(B-V)$, as free
parameters using the extinction law family of \citet{Cardetal89}. For each star we have used as input SEDs
both the CMFGEN grid calculated by \citet{Martetal05} and the TLUSTY OSTAR2002 grid of \citet{LanzHube03}:
however, after running CHORIZOS with the two SED families, we could find no significant differences
between them (not surprising, considering that they both have very similar intrinsic colors for a given
temperature and gravity). Therefore, here we report only on the results for the CMFGEN case. Both 
Pismis 24-1 and Pismis 24-17 have 2MASS $JHK_s$ photometry, though Pismis 24-1 is unresolved. Given the
increased accuracy of the extinction corrections when NIR photometry is included in CHORIZOS, we have done 
two different runs for each object: one with F330W+F435W+F550M+F850LP and another one with those same filters
plus $JHK_s$. Given that Pismis 24-1 is unresolved in 2MASS, for its run with 2MASS we merged the HRC
photometry for 24-1NE and 24-1SW.  

\begin{figure}
\begin{center}
\centerline{\includegraphics*[width=0.97\linewidth]{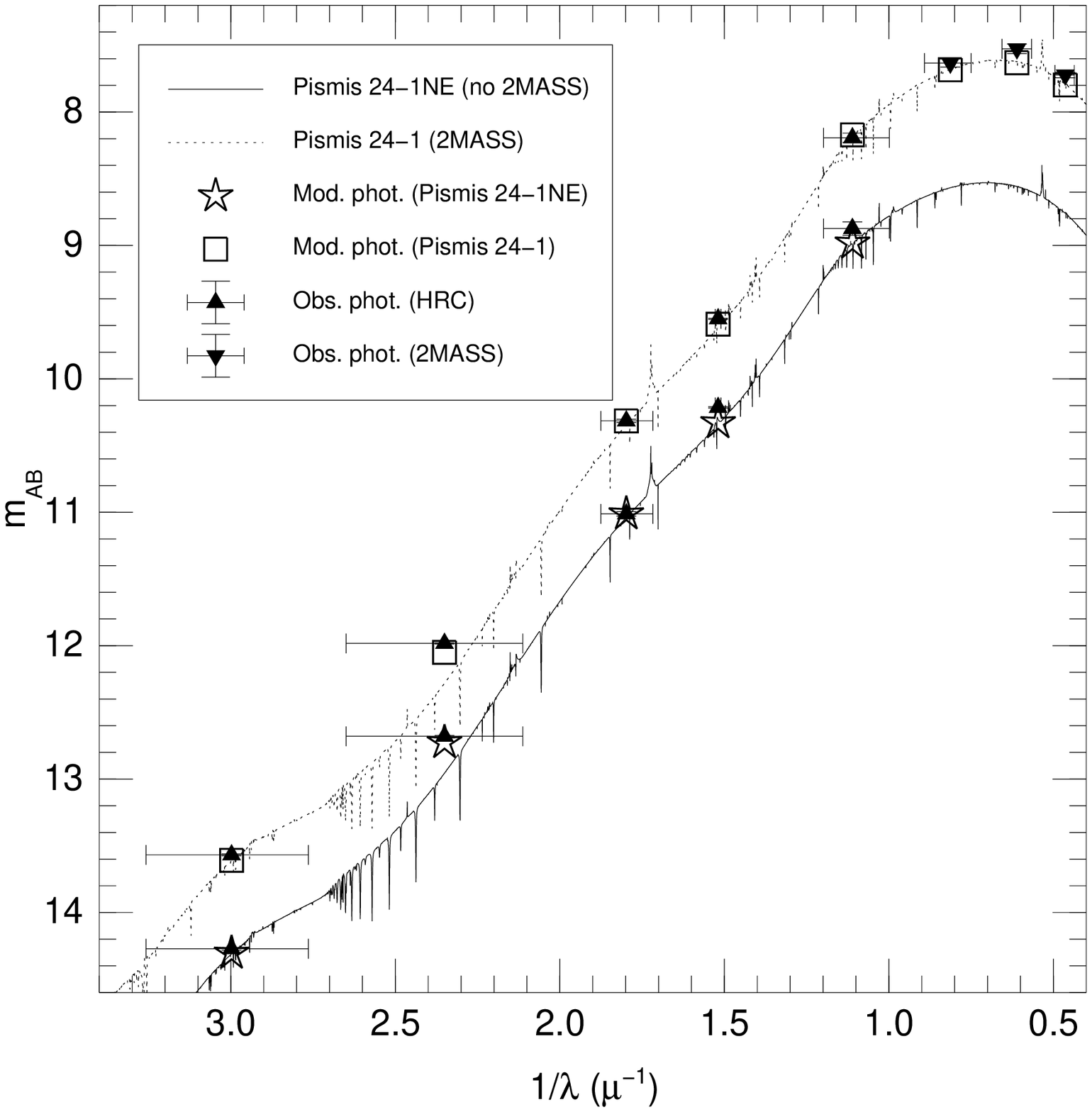}}
\end{center}
\figcaption{Mode SEDs and synthetic photometry for two of the CHORIZOS runs. The observed photometry is
also shown, with the horizontal error bars marking the approximate extent of each of the seven filters
(from left to right, HRC F330W, F435W, F550M, F658N, and F850LP; 2MASS $J$, $H$, and $K_s$) and the vertical
error bars (small in all cases) the photometric uncertainties. The
F658N (H$\alpha$) filter is not included in the CHORIZOS runs but is shown here for comparison: note how the
emission excess is larger for Pismis 24-1NE than for Pismis 24-1. \label{choplot}}
\end{figure}

	The results of the five CHORIZOS runs are shown in Table~\ref{chotable}: the total extinction in
F550M, the extinction law, the absolute visual magnitude (adopting from \citealt{Massetal01} a distance modulus
of $(m-M)_0$ = 12.03$\pm$0.14), the apparent bolometric magnitude (i.e. the observed F550M magnitude with its
extinction and bolometric corrections), the minimum $\chi^2$, and the degrees of freedom of the fit. In
all cases, a single well-defined solution is present in the likelihood $R_{5495}$ vs. $E(4405-5495)$
plots. The mode SED is shown in Fig.~\ref{choplot} for Pismis 24-1NE (without 2MASS) and for Pismis 24-1
(with 2MASS). 

\begin{itemize}
  \item All the runs yield extinctions in F550M close to 6 magnitudes and $R_{5495}$ values close to 3.0. The
	latter are significantly lower than the results found by \citet{Bohietal04}.
  \item Pismis 24-17 is more extinguished than Pismis 24-1 by $\approx$0.4 magnitudes, which is consistent
	with the \citet{Bohietal04} results, though our values are lower overall.
  \item The inclusion of the 2MASS data introduces sginificant changes in the results, increasing the measured 
	F550M extinction by $\approx$0.4 magnitudes and $R_{5495}$ by $\approx$0.15.
  \item The minimum $\chi^2$ values are too high. We attribute this to the \citet{Cardetal89} extinction
	law, which was derived with an older photometric calibration \citep[see][]{Maiz06a} and using
	only moderately-extinguished stars. It also uses a rather unphysical seventh-degree polynomial 
	in the optical region, which is responsible for the long-period oscillations in wavelength visible
	in the reddened SEDs in Fig.~\ref{choplot}, an effect that is unlikely to be real and that
	artificially increases the $\chi^2$ values.
\end{itemize}

	We have used the CHORIZOS results to derive new evolutionary zero-age masses for the three objects in
our sample. The evolutionary tracks have been obtained from \citet{LejeScha01}. Additional uncertainties of 0.1
magnitudes have been added in quadrature to the bolometric magnitudes to take into account the problems with 
the extinction correction and the (small) uncertainty in the effective temperatures. The 2MASS information was included
in the calculations because its presence significantly reduces the systematic errors introduced by the seventh-degree
polynomial used for the extinction law in the optical (in any case, it is always good practice to add NIR data to 
optical photometry when analyzing objects with moderate or high extinction). We obtain nearly
identical values for the three objects: 97$\pm$10 M$_\odot$ (Pismis 24-1NE), 96$\pm$10 M$_\odot$ 
(Pismis 24-1SW), and 92$\pm$9 M$_\odot$ (Pismis 24-17). The real mass for the unresolved components in
Pismis 24-1NE must be, of course, smaller: if the luminosity is equally split and the temperature is the same for both,
the resulting values are of the order of 64 M$_\odot$ for each.  

	As a test, we can also compute the mass of Pismis 24-1 as if it were a single star (i.e. the W02 hypothesis) and we 
obtain\footnote{Note that the evolutionary track with the highest mass is 120 M$_\odot$, so we are forced to extrapolate,
a procedure that should introduce additional uncertainties.} 155$\pm$19 M$_\odot$. This value is much lower than those 
obtained by W02 (210 or 291 M$_\odot$). Given that our value for $M_V$ is very similar to theirs, the difference must arise from 
the different temperature scales used (\citealt{Martetal05}, derived from line-blanketed models, vs. \citealt{Vaccetal96}).
The reduction in temperature for a given spectral type motivated by the change moves the star in an HR diagram towards the lower 
right (the vertical effect is due to the smaller bolometric correction). Such a displacement is almost perpendicular to the 
evolutionary tracks for O stars that are just evolving off the main sequence and, thus, is a very effective way of reducing 
the mass estimate.

	What else could be affecting the masses? Even though Pismis 24 suffers moderate extinction ($A_V\approx 6$ magnitudes),
future improvements in this respect are likely not to be too large, since our analysis includes multiband optical and NIR data 
simultaneously and $A_K$ is only $\approx$ 0.6 magnitudes. Also, our HST optical photometry is compatible with modern ground-based 
results and Tycho did not detect photometric variability, so the possibility of undetected changes (other than those caused by the
known eclipses) seems unlikely. Finally, the distance calculated by \citet{Massetal01} could be wrong: we plan to test that
possibility in the near future by making use of our additional HST data.


\section{Conclusions}

	We have presented HST and ground-based data that clearly resolve Pismis 24-1 into two objects 
separated by 363.86 mas and that indicate the existence of at least a third unresolved component in Pismis
24-1NE. We have also derived zero-age evolutionary masses for the three most massive objects at the core of
Pismis 24-1, with values of 92 to 97 M$_\odot$, one of them corresponding to the
unresolved system Pismis 24-1NE. These values are very large, making the cores of Pismis 24-1 and
Trumpler 14 \citep{Nelaetal04} the two locations within 3 kpc of the Sun with the highest density of very 
massive stars.  However, these masses are not above what it is currently thought to be the stellar upper mass limit.

\begin{acknowledgements}

After this manuscript was submitted, Virpi Niemela, one of the authors, passed away. The rest
of the authors want to dedicate this article to her memory. She was the first one of us to study Pismis 24-1 and also 
the first one to suspect its multiplicity. But, more importantly, she was a wonderful human being to whom a generation of
astronomers is indebted for her insight and her warmth. 

We would like to thank Phil Massey and his team of collaborators for letting us know about the optical
photometric variability prior to publication and David Osip and Mark Phillips for kindly allowing us to
use IMACS during part of an engineering night at the Baade telescope. We also acknowledge the useful comments made
by an anonymous referee. Support for this work was provided by 
(a) the Spanish Government through grant AYA2004-08260-C03; (b) NASA through grant GO-10602 from the 
Space Telescope Science Institute, which is operated by the Association of Universities for Research in
Astronomy Inc., under NASA contract NAS~5-26555; and (c) IALP-CONICET, Argentina.

\end{acknowledgements}

\bibliographystyle{apj}
\bibliography{general}

\end{document}